\documentclass[useAMS,usenatbib,fleqn]{mn2e}
\usepackage[draft]{hyperref}
\usepackage{amssymb}
\usepackage{amsmath}
\usepackage{mathtools}
\usepackage{enumitem}
\usepackage{graphicx}

\usepackage{nicefrac}
\usepackage{bigints}

\usepackage{aas_macros}

\newcommand*{\possis}{\textsc{possis}}
\newcommand*{\altpossis}{{\mdseries\possis}}
          
\title[SN + KN spectra, light curves $\&$ polarization]{\altpossis: predicting spectra, light curves and polarization for multi-dimensional models of supernovae and kilonovae}
\author[M. Bulla]{M.~Bulla,\thanks{E-mail: mattia.bulla@fysik.su.se}\\
Oskar Klein Centre, Department of Physics, Stockholm University, SE 106 91 Stockholm, Sweden
}

\usepackage{color}

\newcommand{\referee}[1]{\textcolor{black}{#1}}
\newcommand{\refereetwo}[1]{\textcolor{black}{#1}}

\begin{document}

\maketitle 

\begin{abstract}
We present \textsc{possis}, a time-dependent three-dimensional Monte Carlo code for modelling radiation transport in supernovae and kilonovae. The code incorporates wavelength- and time-dependent opacities and predicts viewing-angle dependent spectra, light curves and polarization for both idealized and hydrodynamical explosion models. We apply the code to a kilonova model with two distinct ejecta components, one including lanthanide elements with relatively high opacities and the other devoid of lanthanides and characterized by lower opacities. We find that a model with total ejecta mass $M_\mathrm{ej}=0.04\,M_\odot$ and half-opening angle of the lanthanide-rich component $\Phi=30^\circ$ provides a good match to GW\,170817/AT\,2017gfo for orientations near the polar axis (i.e. for a system viewed close to face-on). We then show how crucial is the use of self-consistent multi-dimensional models in place of combining one-dimensional models to infer important parameters such as the ejecta masses. We finally explore the impact of $M_\mathrm{ej}$ and $\Phi$ on the synthetic observables and highlight how the relatively fast computation times of \textsc{possis} make it well-suited to perform parameter-space studies and extract key properties of supernovae and kilonovae. Spectra calculated with \textsc{possis} in this and future studies will be made publicly available.
\end{abstract}

\section{Introduction}
\label{sec:introduction}

The field of time-domain astronomy has witnessed a rapid growth in the past decade thanks to the advent of optical sky surveys, including but not limited to PanSTARRS \citep{Kaiser2010}, the Palomar Transient Factory (PTF, \citealt{Law2009}), the All-sky Automated Survey for Supernovae (ASAS-SN, \citealt{Shappee2014}), the Dark Energy Survey (DES, \citealt{DES2016}), the Asteroid Terrestrial-impact Last Alert System (ATLAS, \citealt{Tonry2018}) and the Zwicky Transient Facility (ZTF, \citealt{Graham2019}). Nowadays, \referee{about five} supernovae (SNe) are discovered every night\footnote{Based on statistics available at \url{https://wis-tns.weizmann.ac.il/stats-maps} for classified supernovae.} and this number is expected to increase significantly when the Large Synoptic Sky Survey \citep[LSST,][]{LSST2009,Ivezi2019} comes online. Current surveys are also well-suited \citep[e.g.][]{Andreoni2019,Goldstein2019} to rapidly scan large regions of the sky to search for electromagnetic counterparts of gravitational-wave events and specifically kilonovae (KNe). 

At the same time, the continuous improvement in computational resources has led to a rapid increase in the available hydrodynamical models for both SNe and KNe. A progress in time-domain astronomy is therefore critically tied to connecting state-of-the-art explosion models with the wealth of available and future observations. Among different techniques, a powerful approach to provide such connection is via radiative transfer calculations, which simulate the propagation of light through an external medium and study the interaction between radiation and matter via absorption and scattering processes. This allows the prediction of synthetic observables -- as light curves, spectra and polarization -- that can then be compared to data to place constraints on models.

Over the past three decades, sophisticated radiative transfer codes have been developed and used to investigate both SNe and KNe \referee{\citep[e.g.][]{Hoeflich1993,Blinnikov1998,Hauschildt1999,Utrobin2004,
Dessart2005,Kasen2006,Kromer2009,Bersten2011,
Jerkstrand2011,Tanaka2013,Frey2013,Wollaeger2013,
Kerzendorf2014,Morozova2015,Ergon2018}}. 
These codes have lead to a better understanding of these phenomena and placed important constraints on the underlying physics. However, simulations performed with some of these codes are typically computationally expensive and thus restricted to sampling only a few realizations of the full parameter space. In addition, some codes work in one dimension and do not capture ejecta inhomogeneities and asymmetries and thus the corresponding viewing-angle dependence of the synthetic observables.

Here, we report on upgrades to the time-dependent multi-dimensional Monte Carlo radiative transfer code \possis~(POlarization Spectral Synthesis In Supernovae)\referee{, originally developed as a test-code in \citet[][see their section~3]{Bulla2015}}. Unlike other radiative transfer codes, \possis~does not solve the radiative transfer equation but rather requires opacities as input. This assumption speeds up the calculation significantly and allows the undertaking of parameter-space studies to constrain key properties of the modelled system. In addition, \possis~works in three dimensions and is thus well-suited to studying intrinsically asymmetric models and predict their observability at different viewing angles.

The paper is organized as follows. We provide an outline of \possis~in Section~\ref{sec:code}, focussing particularly on the new features introduced to the code. We then present a two-component KN model against which we test our code in Section~\ref{sec:testmodel}. We finally show and discuss synthetic observables (spectra, light curves and polarization) for this specific model in Section~\ref{sec:synthobs}, before summarizing in Section~\ref{sec:conclusions}. Spectra computed in this and future works are made available at: \url{https://mattiabulla.wixsite.com/personal/models}.

\section{Outline of the code}
\label{sec:code}

Here we provide a summary of the Monte Carlo radiative transfer code \possis~and outline the new features introduced in this work. \possis~was first presented \referee{as a test-code in \citet[][see their section~3]{Bulla2015}} and used to model polarization of both SNe \citep{inserra2016} and KNe \citep{Bulla2019} at individual time snapshots. The main changes introduced in this work are the energy treatment and a temporal dependence in both opacities and ejecta properties, which then allow us to produce time-dependent spectra (flux and polarization) and broad-band light curves. 

\subsection{Model grid}
\label{sec:grid}

A three-dimensional Cartesian grid is given at some reference time $t_0$, with velocity $v_i$, density $\rho_{i,0}$ and temperature $T_{i,0}$ provided for each grid cell $i$. In the case of KNe, the electron fraction $Y_{\mathrm{e},i}$ is also given. The code assumes homologous expansion, i.e. the velocity $v_i$ in each cell is constant (free expansion) and the corresponding radial coordinate $r_i$ is given by
\begin{equation}
r_i = v_i t
\end{equation}
at any time $t$. The grid is expanded at each time-step $j$, the density is scaled as 
\begin{equation}
\rho_{ij} = \rho_{i,0}\,\bigg( \frac{t_j}{t_0} \bigg)^{-3}
\end{equation}
according to homologous expansion while the temperature is scaled as
\begin{equation}
\label{eq:temp}
T_{ij} = T_{i,0}\,\bigg( \frac{t_j}{t_0} \bigg)^{-\alpha}
\end{equation}
with $\alpha>0$.

\subsection{Opacities}
\label{sec:opac}

\possis~can handle \textit{line} opacity from bound-bound transitions ($\kappa_\mathrm{bb}$) and \textit{continuum} opacity from either electron scattering ($\kappa_\mathrm{es}$), bound-free ($\kappa_\mathrm{bf}$) or free-free ($\kappa_\mathrm{ff}$) absorption. Wavelength-dependent opacities can be given either at each time-step or at a reference time $t_\mathrm{ref}$ together with a function $f_\mathrm{opac}(t)$ describing their temporal evolution. 

Two separate modes can be selected to treat bound-bound opacities. The first mode (sob-mode) treats bound-bound opacities using the Sobolev approximation \citep{Sobolev1960}, in which photons interact with each line at a single frequency and thus at a specific location along their trajectory throughout the ejecta. In the second mode (abs-mode), a polynomial fit to the bound-bound opacity is performed (see e.g. \citealt{inserra2016}) and used together with the bound-free and free-free opacities as representative of a ``pseudo-continuum'' absorption component. The former approach is well-suited to predict spectral features associated to individual line transitions, while the latter allows one to predict a featureless ``pseudo-continuum'' flux level.

\subsection{Creating photon packets}
\label{sec:creating}

A number $N_\mathrm{ph}$ of Monte Carlo quanta are created at any time-step. Each of these quanta is assigned a location $\textbf{x}$ and an initial direction $\textbf{n}$, energy $e$, frequency $\nu$ and normalized Stokes vector $\textbf{s}=(1,q,u)$ \footnote{As in \citealt{Bulla2015} we neglect the Stokes parameter $V$ describing circular polarization.}. Following \citet{Abbott1985} and \citet{Lucy1999}, each Monte Carlo quantum is treated as a packet of identical and indivisible photons (hereafter referred to as packet). As explained below, this implies that the same energy is assigned to all packets and that this energy is kept constant during all the interactions. 

The location $\textbf{x}$ is selected either on a pre-defined photospheric surface or according to the distribution of radioactive material, while the initial direction $\textbf{n}$ is sampled assuming either isotropic emission or constant surface brightness. As mentioned above, packets are treated as identical and carry the same amount of energy throughout the simulation. The total energy from the relevant radioactive decay processes, $E_\mathrm{tot}(t_j)$, is then divided equally among all the packets 
\begin{equation}
e(t_j) = \frac{E_\mathrm{tot}(t_j)\,\epsilon_\mathrm{th}}{ N_\mathrm{ph}}~~,
\end{equation}
where $\epsilon_\mathrm{th}$ is a thermalization efficiency, i.e. we neglect $\gamma$-ray transport and assume that a fraction $\epsilon_\mathrm{th}$ of $E_\mathrm{tot}(t_j)$ is deposited and made available for ultraviolet-optical-infrared radiation. The initial frequency $\nu$ is chosen by sampling the thermal emissivity
\begin{equation}
\label{eq:thermal}
S(\nu) = B(\nu,T)\,\kappa_\mathrm{tot}(\nu)~~,
\end{equation}
where $\kappa_\mathrm{tot}(\nu)$ is the total opacity and $B(\nu,T)$ is the Planck function at temperature $T$. Finally, packets are created unpolarized, i.e. their normalized Stokes vector is set to $\textbf{s} = (1,0,0)$.

\begin{figure}
\begin{center}
\includegraphics[width=1\columnwidth]{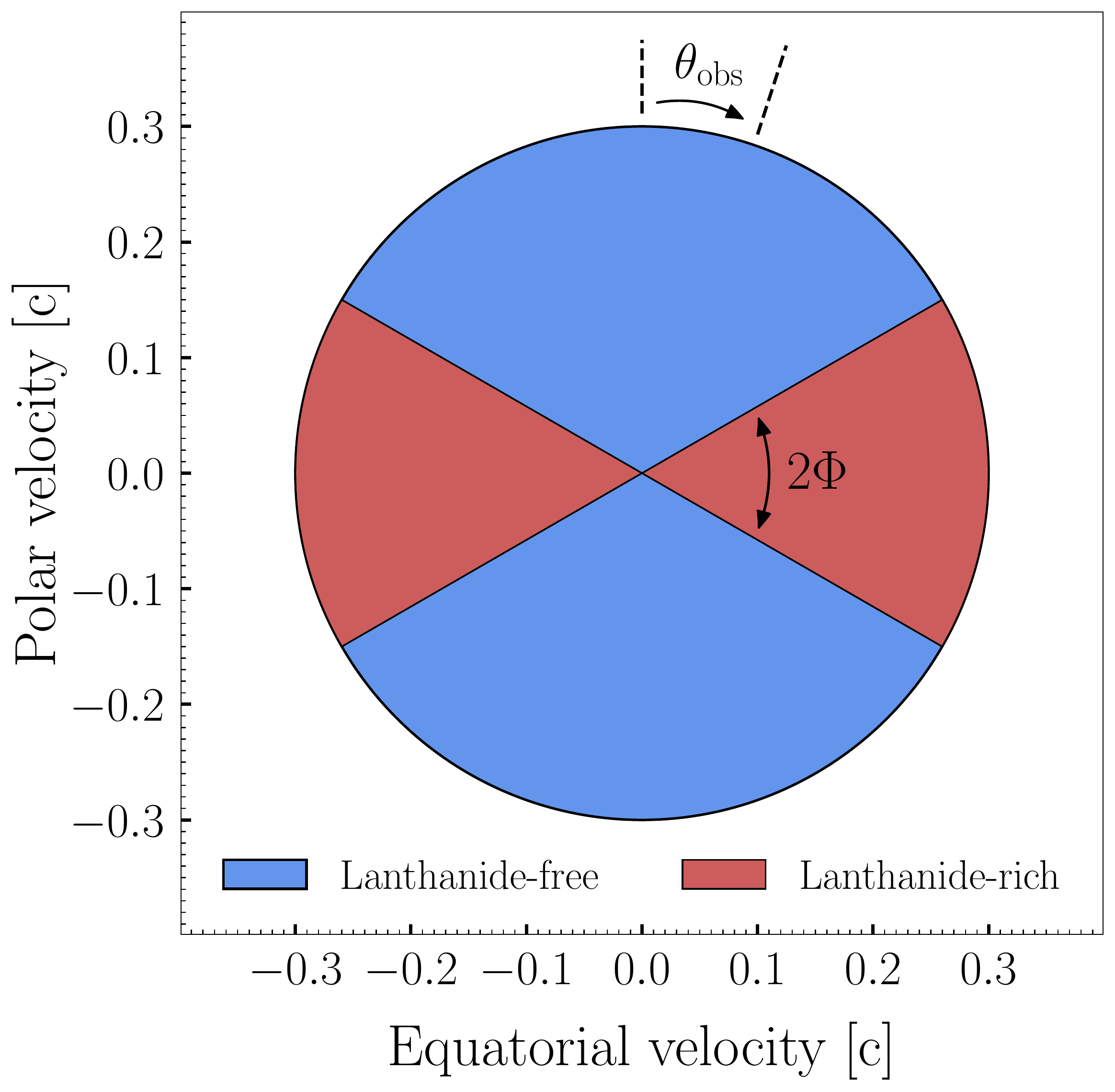}
\caption{A meridional cross-section of the two-component kilonova model adopted in this study. A ``lanthanide-rich'' component is distributed around the merger plane (with half-opening angle $\Phi$) and characterized by high opacities from lanthanides (red region). A ``lanthanide-free'' component is distributed at higher latitudes and characterized by lower opacities (blue region). Synthetic observables are calculated for different viewing angles $\Theta_\mathrm{obs}$.}
\label{fig:geo}
\end{center}
\end{figure}

\subsection{Propagating photon packets}
\label{sec:propagating}

Each packet is propagated throughout the ejecta until it interacts with matter. The propagation of a packet is performed in the rest frame, while interactions are treated in the comoving frame. This involves transforming properties like the direction of propagation and frequency from rest frame to comoving frame (and viceversa) every time an interaction with matter occurs (see \citealt{Bulla2015} for details).

Which event occurs is chosen depending on the mode selected to treat bound-bound opacity (see Section~\ref{sec:opac}). In the sob-mode, the procedure outlined in \citet{Bulla2015} is adopted to select whether a line or continuum interaction occurs. In the abs-mode, instead, a continuum event is selected. When a continuum interaction is selected in either modes, a random number $\xi$ is drawn from a uniform distribution over the interval $[0,1)$ to determine the nature of the event. Specifically, electron scattering is selected if 
\begin{equation}
\xi<\frac{\kappa_\mathrm{es}}{\kappa_\mathrm{es}+\kappa_{abs}}~~, 
\end{equation}
where $\kappa_\mathrm{abs}=\kappa_\mathrm{bf}+\kappa_\mathrm{ff}$ in the sob-mode while $\kappa_\mathrm{abs}=\kappa_\mathrm{bb}+\kappa_\mathrm{bf}+\kappa_\mathrm{ff}$ in the abs-mode. Continuum absorption is chosen otherwise.

Upon interaction, the properties of a packet are updated according to the specific event that occurred. In the case of electron scattering, a new direction and Stokes vector are calculated according to the scattering angles randomly selected (see \citealt{Bulla2015}) while the frequency of the packet is kept unchanged. In the cases where bound-bound, bound-free or free-free opacity is selected, the packet is instead re-emitted isotropically, with no polarization and with a new frequency. The latter is calculated using the ``two-level atom'' (TLA) approach described by \citet{Kasen2006}, in which a packet can be re-emitted either at the same frequency or at a new frequency sampled from the thermal emissivity of the given cell (see equation~\ref{eq:thermal}). The probability of redistribution is controlled by the redistribution parameter $\epsilon$, which is set to $\epsilon=0.9$ following \citet{Magee2018}.

The procedure described in this Section is repeated until the packet leaves the computational boundary.

\begin{figure}
\begin{center}
\includegraphics[width=1\columnwidth]{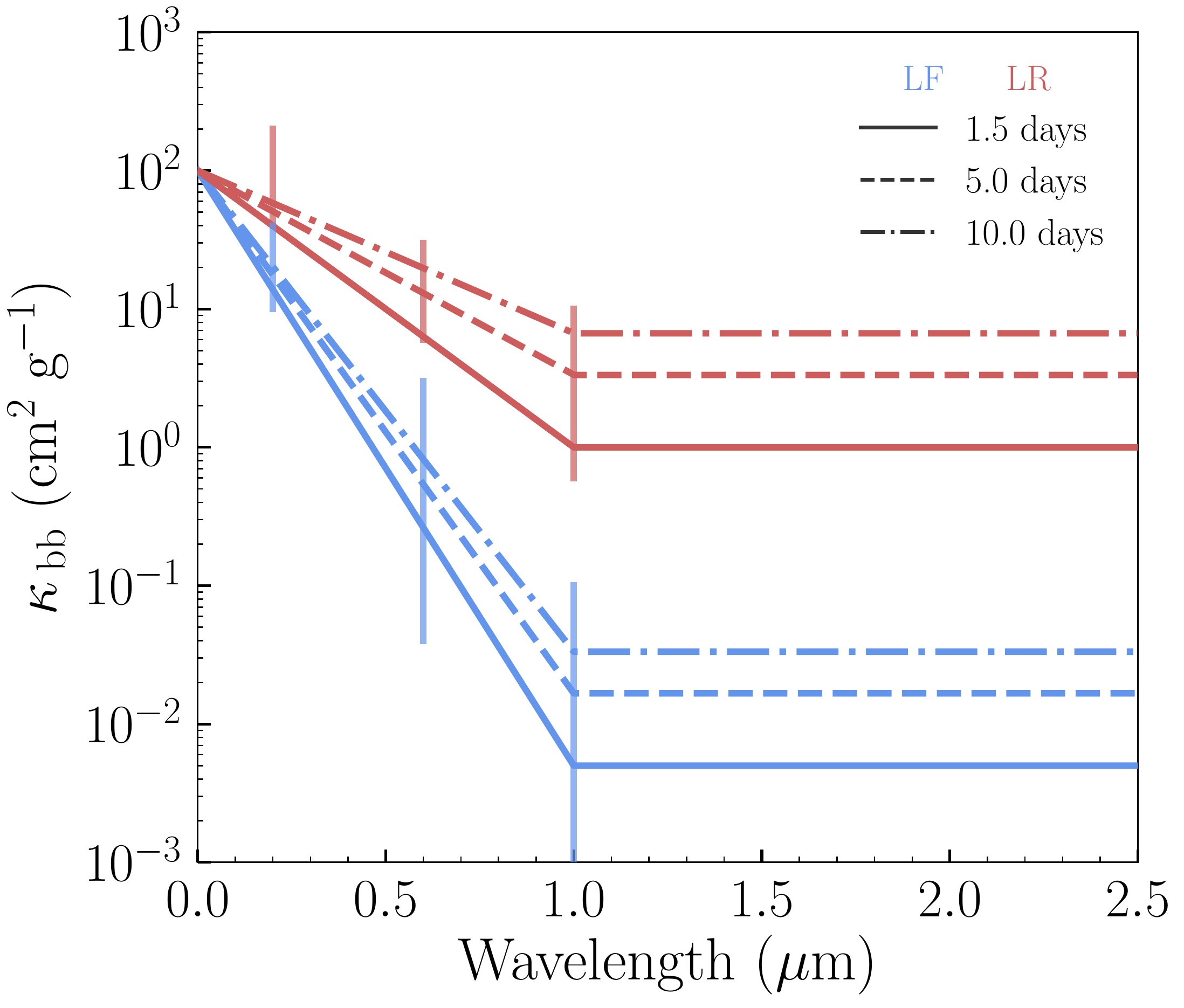}
\caption{Bound-bound line opacities $\kappa_\mathrm{bb}$ adopted in this study for the lanthanide-free (blue) and lanthanide-rich (red) component. Opacities are shown at three different epochs: 1.5 (solid), 5 (dashed) and 10 (dot-dashed) days after the merger. \referee{Vertical lines show the range of opacities at 1~d and 0.2, 0.5 and 1 $\mu$m spanned by models with $Y_\mathrm{e}\leq 0.25$ (lanthanide-rich, red) and $Y_\mathrm{e}>0.25$ (lanthanide-free, blue) from state-of-the-art calculations by \citet[][see their figure 9]{Tanaka2019}}.}
\label{fig:opac}
\end{center}
\end{figure}

\subsection{Collecting photon packets}
\label{sec:collecting}

\begin{figure*}
\begin{center}
\includegraphics[width=1\textwidth]{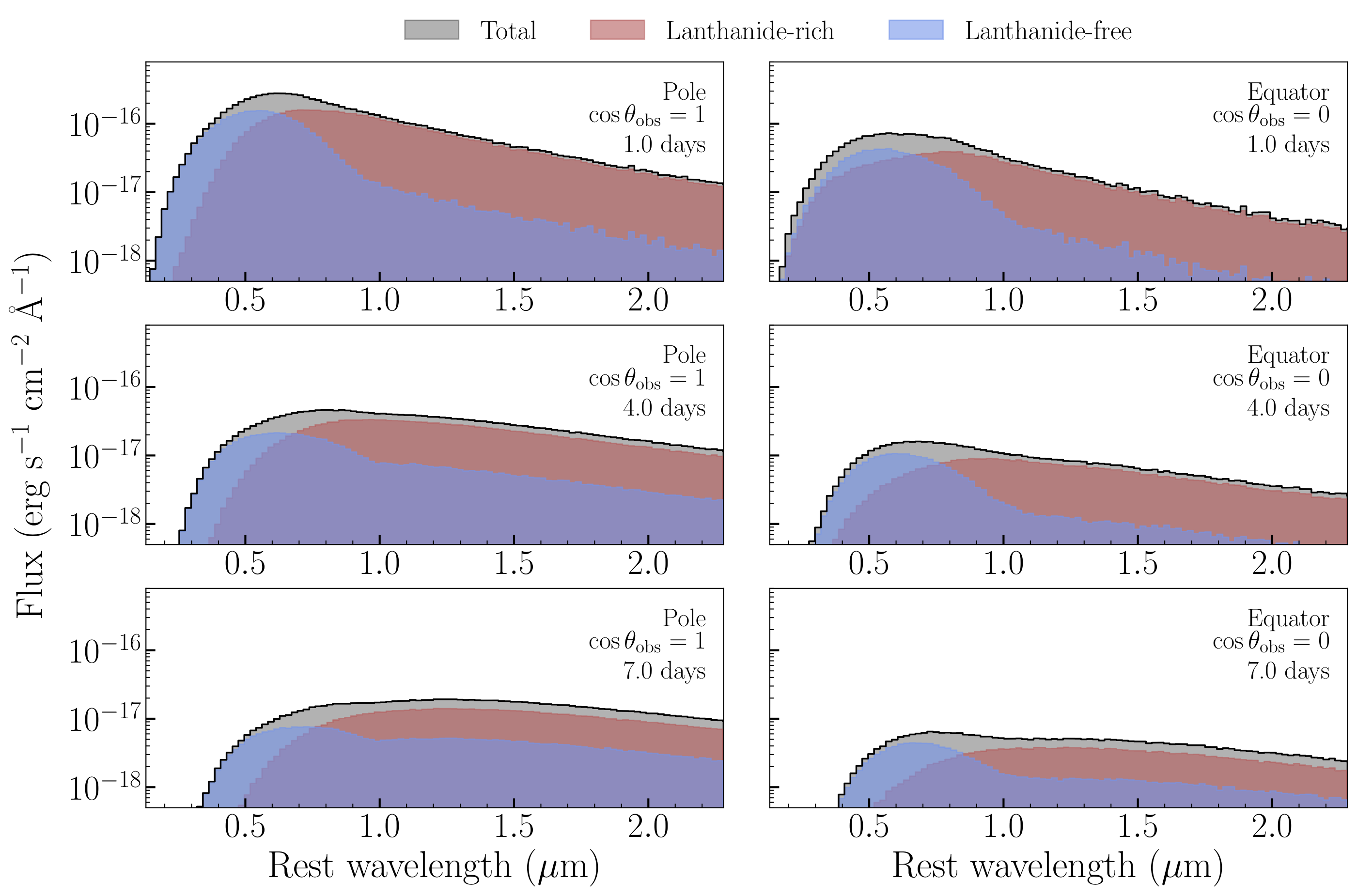}
\caption{Spectral energy distributions (SEDs) of the nsns\underline{\hspace{0.15cm}}mej0.04\underline{\hspace{0.15cm}}phi30 model for an observer along the polar axis ($\cos\theta_\mathrm{obs}=1$, left panels) and one in the equatorial plane ($\cos\theta_\mathrm{obs}=0$, right panels). SEDs are shown at 1.0 (top), 4.0 (middle) and 7.0 (bottom) days after the merger. In each panel, shaded area highlight contribution from photon packets that have their last interaction in the lanthanide-free (blue) and lanthanide-rich (red) component. Fluxes are scaled at 40 Mpc, i.e. at the distance inferred for AT\,2017gfo \citep{Freedman2001,LigoGCN2017}. }
\label{fig:speccontr}
\end{center}
\end{figure*}

Two different approaches are used simultaneously by \possis~to predict synthetic observables: a direct counting technique (DCT) and an event-based technique (EBT). In the former approach -- typically adopted in Monte Carlo radiative transfer codes -- packets escaping the computational boundary are collected in different angular bins according to their final directions $\textbf{n}$. The resulting spectra are then computed as
\begin{equation}
\begin{pmatrix} I \\ Q \\ U \end{pmatrix} = \sum \frac{e}{\Delta t~ \Delta\nu~4\pi r^2 }~\textbf{s}_\text{f} ~~,
\end{equation}
where $r$ is the distance between the observer and the system and the sum is performed over all the packets arriving to the observer with a final Stokes vector $\textbf{s}_\mathrm{f}$ in the time interval [$t-\Delta t/2$, $t-\Delta t/2$] and frequency range [$\nu-\Delta \nu/2$, $\nu-\Delta \nu/2$]. $I$  is used to calculate flux spectra and light curves, while all three Stokes parameters $I$, $Q$ and $U$ are used to compute polarization spectra.

In the EBT, virtual packets are created every time a Monte Carlo packet interact with matter. Virtual packets are then sent directly to $N_\mathrm{obs}$ specific observer orientations defined at the start of the simulation, with energy, frequency and Stokes vector equal to those calculated for the real packets after the interaction (see Section~\ref{sec:propagating}). Virtual packets are weighted according to the probability of reaching the observer, which takes into account (i) the probability per unit solid angle $dP/d\Omega|_\text{EBT}$ of being scattered in the observer direction (see equation~16 of \citealt{Bulla2015}) and (ii) the probability of reaching the computational boundary (and thus the observer) without further interaction, $e^{-\tau_\mathrm{esc}}$ (where $\tau_\mathrm{esc}$ is the optical depth to the boundary, see equation~17 of \citealt{Bulla2015}). To speed up the calculations, we follow \cite{Bulla2015} and neglect virtual packets with  $\tau_\mathrm{esc}> \tau^\mathrm{max}_\mathrm{esc}=10$. Synthetic observables can then be calculated for the pre-defined $N_\mathrm{obs}$ observer viewing angles. In particular, spectra are computed as 
\begin{equation}
\begin{pmatrix} I \\ Q \\ U \end{pmatrix} = \sum \frac{e}{\Delta t~ \Delta\nu~r^2 }~s_\text{f} \cdot\bigg(\frac{dP}{d\Omega}\bigg|_\text{EBT} ~e^{-\tau_\text{esc}}\bigg)~~.
\end{equation}
for each viewing angle. Compared to the DCT, the EBT allows one to calculate synthetic observables with much smaller Monte Carlo noise levels and avoids the need to average contributions from different angles in the same angular bin \citep{Bulla2015}.

\section{A test model for kilonovae}
\label{sec:testmodel}

\referee{As mentioned in Section~\ref{sec:code}, our radiative transfer code \possis~is well-suited to calculate synthetic observables for both SN and KN models. In this study, \referee{however, we choose to} test the code \possis~by computing spectra, light curves and polarization for the two-component KN model of \citet{Bulla2019}}. 

Fig.~\ref{fig:geo} shows a meridional cross-section of the adopted ejecta morphology. The model is axially symmetric and characterized by two distinct ejecta components: (i) a ``lanthanide-rich'' component distributed around the merger plane with half-opening angle $\Phi$ and (ii) a ``lanthanide-free'' component distributed at higher latitudes. Broadly speaking, these two components can be thought of as the dynamical ejecta and lanthanide-free post-merger ejecta (disk wind), respectively.

We adopt the main source of opacities in KNe, i.e. electron scattering and bound-bound opacities. We fix opacities at a reference time $t_\mathrm{ref}=1.5$~d after the merger and use simple prescriptions for their time-evolution. Choices of the opacities are guided by numerical simulations from \referee{\citet{Tanaka2019}}, with bound-bound opacities treated in the abs-mode (see Section~\ref{sec:opac}). As shown in Fig.~\ref{fig:opac}, we adopt a power-law dependence of bound-bound opacities on wavelength below 1~$\mu$m while we choose the same value of $\kappa_\mathrm{bb}$ at longer wavelengths. Specifically, electron scattering opacities are taken as 
\begin{equation}
\kappa_\mathrm{es}^\mathrm{lf}=\kappa_\mathrm{es}^\mathrm{lr}=0.01\,\bigg(\frac{t}{t_\mathrm{ref}}\bigg)^{-\gamma}~\mathrm{cm}^2\,\mathrm{g}^{-1} 
\label{eq:kappaes}
\end{equation}
while bound-bound opacities controlled by their value at 1~$\mu$m, which is allowed to vary as
\begin{equation}
\kappa_\mathrm{bb}^\mathrm{lf}[1\mu\mathrm{m}]=5\times10^{-3}\,\bigg(\frac{t}{t_\mathrm{ref}}\bigg)^{\gamma}~\mathrm{cm}^2\,\mathrm{g}^{-1} 
\label{eq:kappabb}
\end{equation}
for the lanthanide-free component and as
\begin{equation}
\kappa_\mathrm{bb}^\mathrm{lr}[1\mu\mathrm{m}]=1.0\,\bigg(\frac{t}{t_\mathrm{ref}}\bigg)^{\gamma}~\mathrm{cm}^2\,\mathrm{g}^{-1} 
\end{equation}
for the lanthanide-rich component. Models with different choices of $\gamma$ are calculated, but in this study we will focus on results with $\gamma=1$, a value that is found to give good fits to the AT\,2017gfo data (see Section~\ref{sec:synthobs}).

We adopt a power-law density profile, i.e. the density in each cell $i$ is initialized as
\begin{equation}
\rho_{i,0} = A\,r_i^{-\beta}~~,
\end{equation} 
where the power-law index is set to $\beta=3$ (in line with predictions from hydrodynamical calculations, \citealt{Hotokezaka2013,Tanaka2013}) and the scaling constant $A$ derived to give a desired ejecta mass $M_\mathrm{ej}$. The temperature is assumed to be uniform throughout the ejecta, its initial value set to $T_{i,0}=5000$~K and the power-law index describing the temporal evolution (see equation~\ref{eq:temp}) fixed to $\alpha=0.4$. The total energy $E_\mathrm{tot}(t_j)$ is calculated from the nuclear-heating rates of \citet[][see their equation~4]{Korobkin2012} and a thermalization factor $\epsilon_\mathrm{th}=0.5$ is assumed. Packets are created according to the distribution of radioactive materials and assuming isotropic emission.

Flux spectra and light curves presented in this work are extracted from simulations using $N_\mathrm{ph}=10^6$, while polarization spectra are from higher signal-to-noise calculations with $N_\mathrm{ph}=2\times10^7$. Observables are computed between 0.5 and 15~d after the merger ($\Delta t=0.5$~d) and in the wavelength range $0.1-2.3\,\mu$m ($\Delta \lambda=0.022\,\mu$m). The EBT approach is adopted and $N_\mathrm{obs}=11$ viewing angles are taken from pole ($\Theta_\mathrm{obs}=0$) to equator ($\Theta_\mathrm{obs}=\pi/2$) equally-spaced in cosine, i.e. $\Delta(\cos\Theta)=0.1$.

\referee{We will focus most of the discussion on a fiducial model with \mbox{$M_\mathrm{ej}=0.04\,M_\odot$} and \mbox{$\Phi=30^\circ$} (denoted as nsns\underline{\hspace{0.15cm}}mej0.04\underline{\hspace{0.15cm}}phi30) while we explore the impact of these two parameters on the light curves in Section~\ref{sec:lc}. The fiducial model is characterized by an ejecta mass of $M_\mathrm{ej}^\mathrm{lr}=0.016\,M_\odot$ in the lanthanide-rich component and an ejecta mass of $M_\mathrm{ej}^\mathrm{lf}=0.024\,M_\odot$ in the lanthanide-free component.}

\section{Synthetic observables}
\label{sec:synthobs}

Here, we present viewing-angle dependent synthetic observables calculated for the model described in Section~\ref{sec:testmodel}. We show spectral energy distributions (SEDs) in Section~\ref{sec:seds}, broad-band light curves in Section~\ref{sec:lc} and polarization spectra in Section~\ref{sec:polarization}. 

\subsection{Spectral energy distribution}
\label{sec:seds}

\begin{figure}
\begin{center}
\includegraphics[width=1\columnwidth]{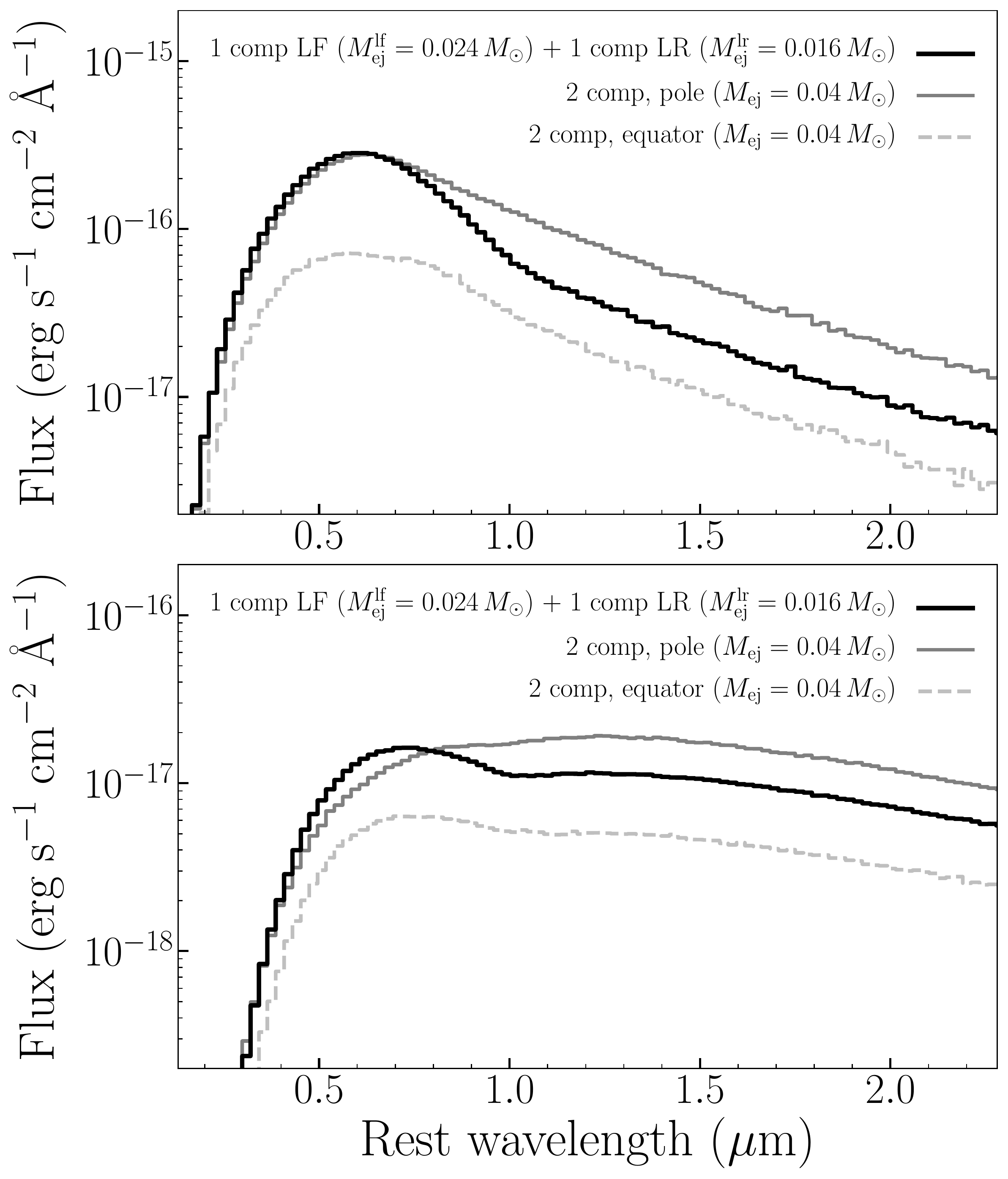}
\caption{Spectral comparison between the two-component model of Fig.~\ref{fig:speccontr} (solid grey, pole; dashed grey, equator) and the sum of a one-component lanthanide-free model with a one-component lanthanide-rich model (black line). The lanthanide-free model ($\Phi=0^\circ$) has a mass equal to that in the lanthanide-free region of the two-component model, \mbox{$M_\mathrm{ej}^\mathrm{lf}=0.024\,M_\odot$}, while the lanthanide-rich model ($\Phi=90^\circ$) to that in the lanthanide-rich region of the two-component model, \mbox{$M_\mathrm{ej}^\mathrm{lr}=0.016\,M_\odot$}. Spectra are shown at 1~d (upper panel) and 7~d (lower panel) after the merger. Fluxes are scaled at 40 Mpc, i.e. at the distance inferred for AT\,2017gfo \citep{Freedman2001,LigoGCN2017}. The comparison highlights how summing one-component models gives incorrect results.  }
\label{fig:spec_1c2c}
\end{center}
\end{figure}

\begin{figure*}
\begin{center}
\includegraphics[width=1\textwidth]{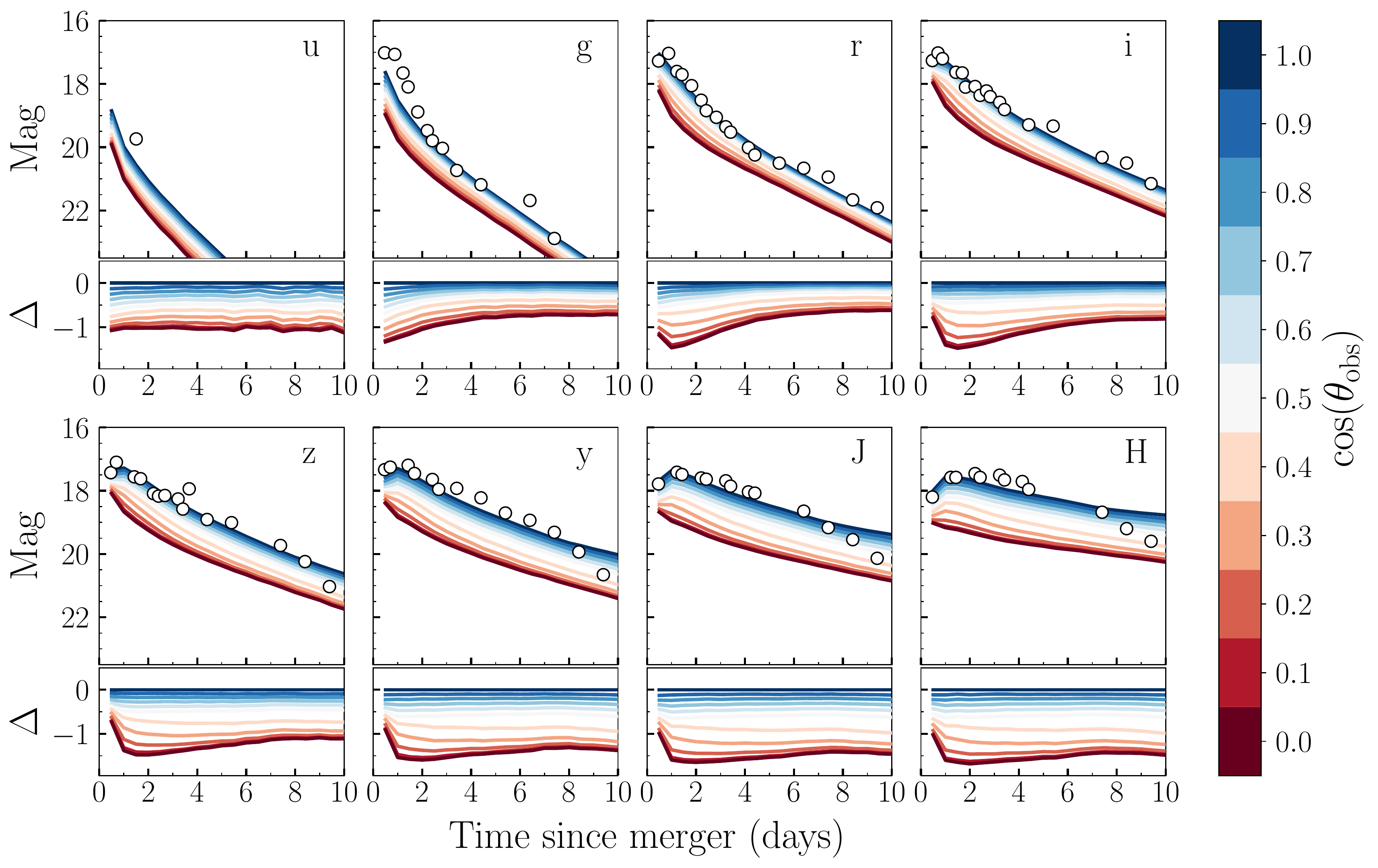}
\caption{Broad-band (ugrizyJH) light curves of the nsns\underline{\hspace{0.15cm}}mej0.04\underline{\hspace{0.15cm}}phi30 model. Light curves are shown for $N_\mathrm{obs}=11$ different viewing angles from equator (dark red, edge-on, $\cos\theta_\mathrm{obs}=0$) to pole (dark blue, face-on, $\cos\theta_\mathrm{obs}=1$). For each filter, a sub-panel shows the difference $\Delta$ between a given viewing angle and the polar direction. Photometry of AT\,2017gfo is \refereetwo{corrected for Milky Way extinction adopting $E(B-V)=0.105$~mag \citep{Schlafly2011} and} shown with open circles in each panel, while models are scaled at 40~Mpc (i.e. at the distance inferred for GW\,170817/AT\,2017gfo, \citealt{Freedman2001,LigoGCN2017}). \refereetwo{Host extinction is suggested to be low \citep[e.g.][]{Pian2017} and thus neglected here.} }
\label{fig:lcfiducial}
\end{center}
\end{figure*}

\begin{figure*}
\begin{center}
\includegraphics[width=\textwidth]{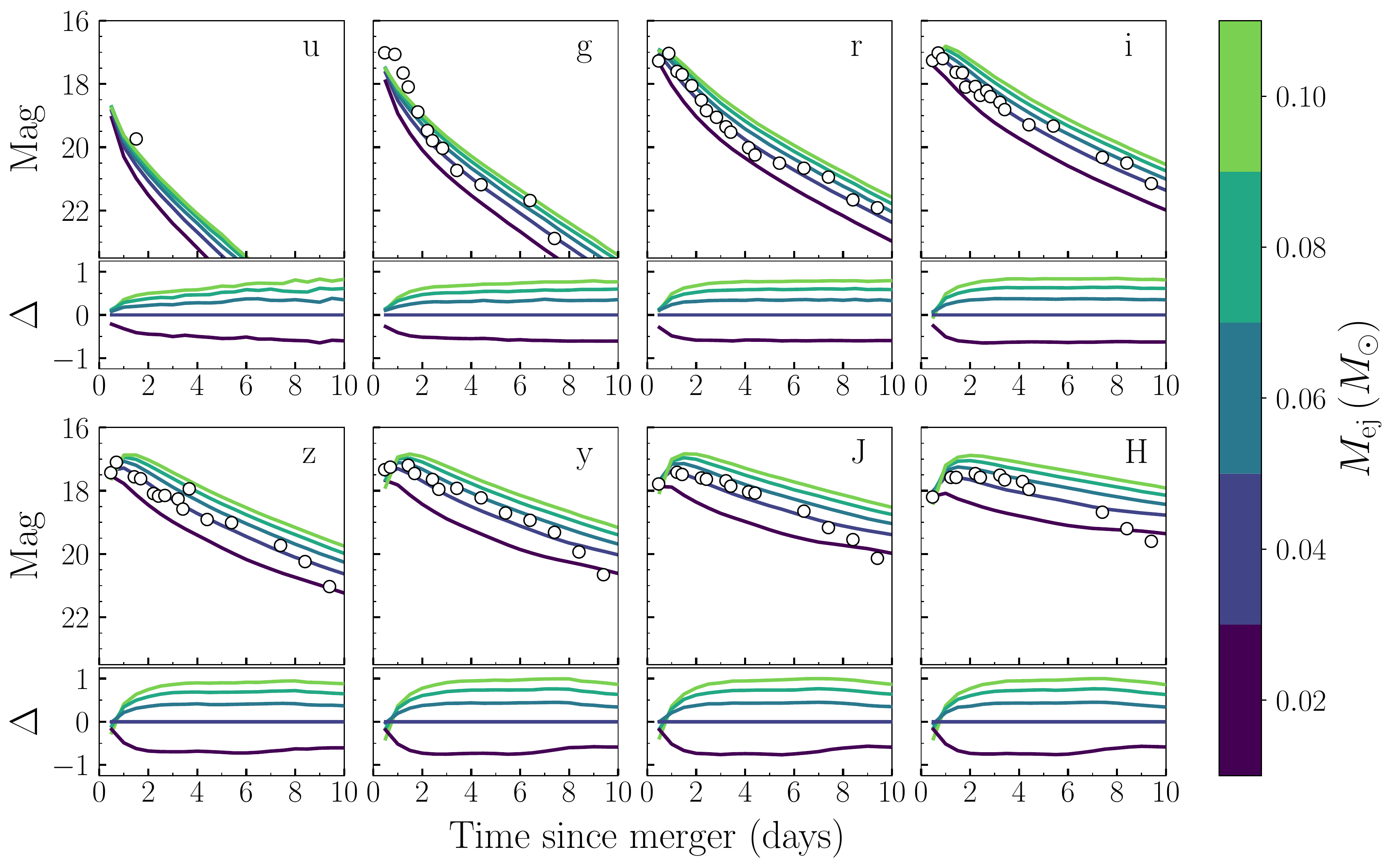}
\includegraphics[width=\textwidth]{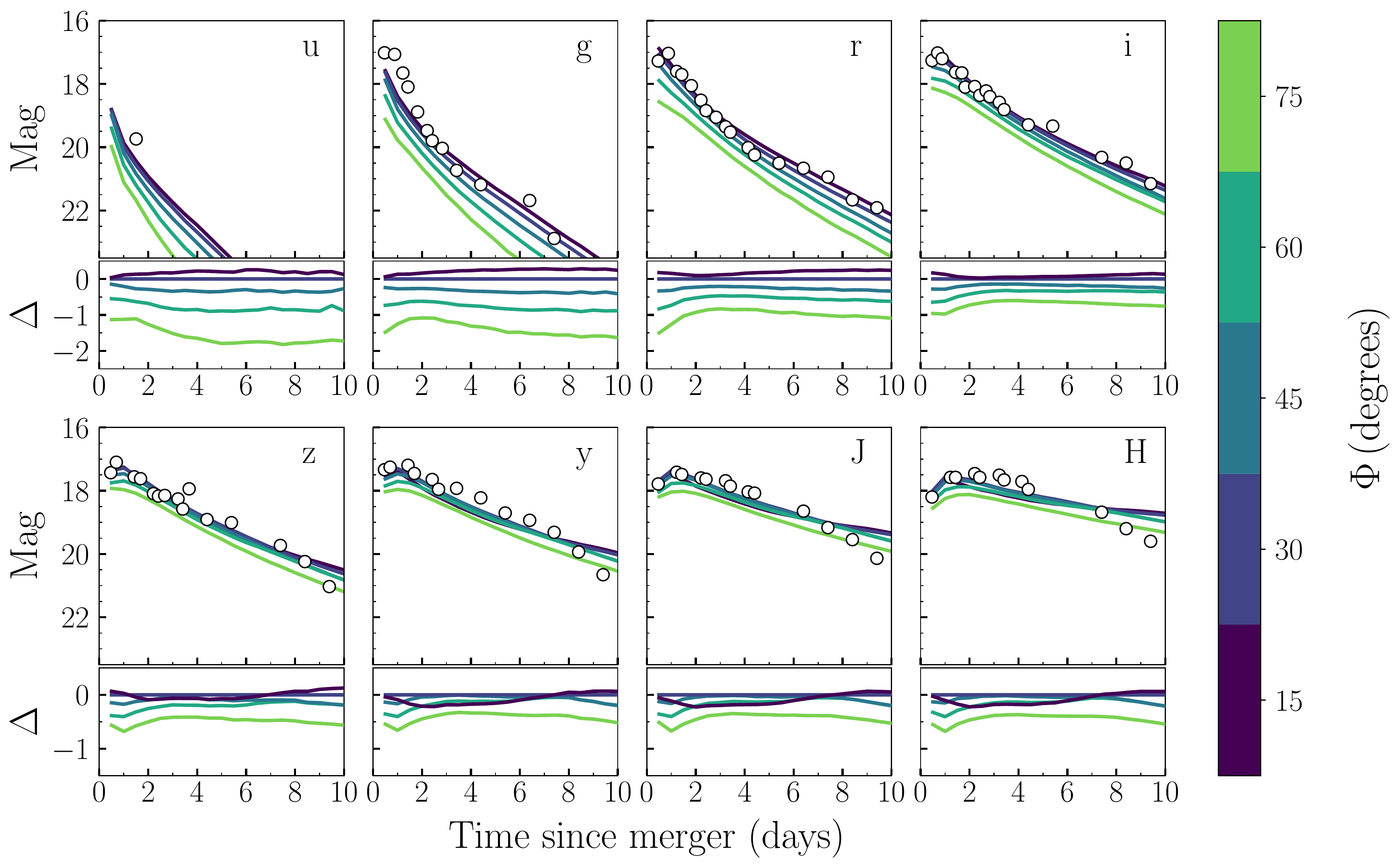}
\caption{Same as Fig.~\ref{fig:lcfiducial} but for a polar viewing angle (face-on, $\cos\theta_\mathrm{obs}=1$) and different ejecta masses $M_\mathrm{ej}$ (upper panels) and half-opening angle of the lanthanide-rich region $\Phi$ (lower panels). Upper panels assume $\Phi=30^\circ$, while bottom panels $M_\mathrm{ej}=0.04~M_\odot$. }
\label{fig:lcdep}
\end{center}
\end{figure*}

SEDs in the first week after the merger are shown in Fig.~\ref{fig:speccontr} for the nsns\underline{\hspace{0.15cm}}mej0.04\underline{\hspace{0.15cm}}phi30 model seen from two different orientations: one looking at the system face-on ($\cos\theta_\mathrm{obs}=1$, left panels) and one edge-on ($\cos\theta_\mathrm{obs}=0$, right panels). At all wavelengths, SEDs are fainter when the system is viewed edge-on compared to face-on. This is a direct consequence of the higher opacities (Section~\ref{sec:testmodel}) and then more severe line-blocking that packets experience trying to escape the ejecta through equatorial rather than polar regions.

Each panel of Fig.~\ref{fig:speccontr} shows the contribution to the total flux of packets coming from the two distinct components. Packets travelling into the lanthanide-rich region are very likely to interact multiple times with lines and thus to be first absorbed and then re-emitted at longer wavelengths. Hence, flux coming from the lanthanide-rich region emerges preferentially in the infrared. In contrast, interactions with lines occur less frequently for packets travelling in the lanthanide-free component. Hence, flux coming from the lanthanide-free region emerges preferentially in the optical, while the infrared re-processed flux is roughly an order of magnitude smaller compared to that from the lanthanide-rich region. 

The re-processing mechanism described above is also time-dependent. Packets interacting multiple times with lines typically take longer to diffuse out and to finally escape the ejecta. This leads to a clear evolution from an SED peaking in the optical at early times (1 d after the merger) to an SED peaking in the infrared at later times (7 d after the merger). For both viewing angles, this is highlighted by the relative increase of infrared compared to optical flux in the lanthanide-free component. The predicted time-evolution accounts for the transition from a so-called ``blue'' KN to a ``red'' KN that was observed in AT\,2017gfo \citep[e.g.][]{Cowperthwaite2017,Pian2017,Kasliwal2017,Shappee2017,Smartt2017}.

Fig.~\ref{fig:spec_1c2c} shows the sum of a one-component lanthanide-free model ($\Phi=0^\circ$) with a one-component lanthanide-rich model ($\Phi=90^\circ$), in the following referred to as the 1cLF+1cLR model. Combinations of this sort have been reported in the literature to infer the presence of two ejecta components in GW\,170817/AT\,2017gfo and to extract their ejecta masses (e.g. \citealt{Kasen2017}, \citealt{Chornock2017}, \citealt{Kilpatrick2017} and \citealt{Nicholl2017}). Fig.~\ref{fig:spec_1c2c} highlights how SEDs thus calculated are different from those computed with our self-consistent two-component model at different times. At 1~d after the merger (upper panel), the 1cLF+1cLR model has nearly the same brightness as the face-on two-component model ($\cos\theta_\mathrm{obs}=1$) in the optical, but it is a factor of $\sim2$ fainter in the infrared. At later epochs (e.g. 7~d, lower panel) the difference is even stronger, with the 1cLF+1cLR model inconsistent with any viewing angle of the two-component model. Based on this comparison, we argue against combining one-component models with different compositions to interpret KN data and infer key parameters as e.g. ejecta masses $M_\mathrm{ej}^\mathrm{lf}$ and $M_\mathrm{ej}^\mathrm{lr}$. 

\subsection{Broad-band light curves}
\label{sec:lc}

Fig.~\ref{fig:lcfiducial} shows broad-band light curves predicted for the nsns\underline{\hspace{0.15cm}}mej0.04\underline{\hspace{0.15cm}}phi30 model. In particular, ugrizyJH light curves are shown for $N_\mathrm{obs}=11$ viewing angles against data collected in the same bands for AT\,2017gfo \citep{Andreoni2017,Arcavi2017,Chornock2017,
Cowperthwaite2017,Drout2017,Evans2017,
Kasliwal2017,Pian2017,Smartt2017,Tanvir2017,
Troja2017,Utsumi2017,Valenti2017}. 

Owing to the difference in SEDs at different orientations (see Section~\ref{sec:seds}), the viewing-angle dependence of the light curves is also quite strong. Specifically, an observer in the merger plane (system viewed edge-on, $\cos\theta_\mathrm{obs}=0$) would see a KN $\sim$~1$-$1.5~mag fainter than an observer along the polar axis (system viewed face-on, $\cos\theta_\mathrm{obs}=1$) depending on the specific filters. The small panels in Fig.~\ref{fig:lcfiducial} highlight a viewing-angle dependence in the light-curve shape as well. In particular, the magnitude difference between a face-on and edge-on KN tends to decrease with time. This is a direct consequence of the different diffusion time-scales at different orientations, with photons escaping the ejecta near the equator interacting multiple times within the lanthanide-rich component and thus arriving to the observer later (see also discussion in Section~\ref{sec:seds}). Because line opacities are higher at optical rather than infrared wavelengths, this effect is most evident in the gri filters, highlighting the importance of optical observations to constrain the inclination of future KN events.

After scaling the model fluxes to the distance inferred for AT\,2017gfo \citep[40\,Mpc,][]{Freedman2001,LigoGCN2017}, we find a better agreement with data for viewing angles close to the polar axis (blue lines in Fig.~\ref{fig:lcfiducial}). This is consistent with previous findings suggesting that AT\,2017gfo was observed at 15$^\circ\lesssim\theta_\mathrm{obs}\lesssim30^\circ$ ($0.87\lesssim\cos\theta_\mathrm{obs}\lesssim0.97$) from the polar axis \citep{Abbott2017,Pian2017,Troja2017,Finstad2018,
Mandel2018,Mooley2018}. The good agreement is especially true at bluer wavelengths (ugri). For redder filters (zyJH), models are consistent with data in the first week after the merger whereas they tend to decline more slowly than observed at later epochs. This discrepancy points to an incorrect assumption for the time-dependence of opacities in the near-infrared (see discussion in Section~\ref{sec:testmodel}).

The impact of the ejecta mass $M_\mathrm{ej}$ on the predicted light curves is shown in the top panels of Fig.~\ref{fig:lcdep} for an observer looking at the system from the polar axis (face-on, $\cos\theta_\mathrm{obs}=1$). A larger $M_\mathrm{ej}$ translates into a brighter KN in all filters following the increase in the amount of radioactive material (i.e. energy budget). At the same time, however, higher ejecta masses provide larger opacities to radiation. As shown in Fig.~\ref{fig:lcdep}, this has two effects when moving to increasingly larger masses: the increase in brightness tends to plateau and the light curves tend to peak later (due to increasingly larger diffusion time-scales, see especially near-infrared bands).

The impact of the half-opening angle $\Phi$ on the predicted light curves is shown in the bottom panels of Fig.~\ref{fig:lcdep} for an observer along the polar axis (face-on, $\cos\theta_\mathrm{obs}=1$). For the same total mass $M_\mathrm{ej}$, varying the $\Phi$ value has the effect of changing the relative fraction of mass in one compared to the other ejecta component, i.e. $M_\mathrm{ej}^\mathrm{lf}$ vs $M_\mathrm{ej}^\mathrm{lr}$. At bluer wavelengths, both components contribute to the spectrum (see Fig.~\ref{fig:speccontr}). Reducing $\Phi$ leads to smaller opacities from the lanthanide-rich region and a larger flux contribution from the lanthanide-free component, effects which combine to give brighter KNe. Redder wavelengths, instead, are dominated by flux coming from the lanthanide-rich region (Fig.~\ref{fig:speccontr}). Initially, reducing $\Phi$ decreases the opacities from the lanthanide-rich component, thus leading to brighter KNe in the infrared. This increase in brightness is seen when lowering $\Phi$ from 75 to 30$^\circ$. Reducing $\Phi$ from 30 to 15$^\circ$, however, leads to a fainter KN in the infrared following a decrease in $M_\mathrm{ej}^\mathrm{lr}$ and thus in the energy budget from the lanthanide-rich region.

\subsection{Polarization}
\label{sec:polarization}

\begin{figure}
\begin{center}
\includegraphics[width=\columnwidth]{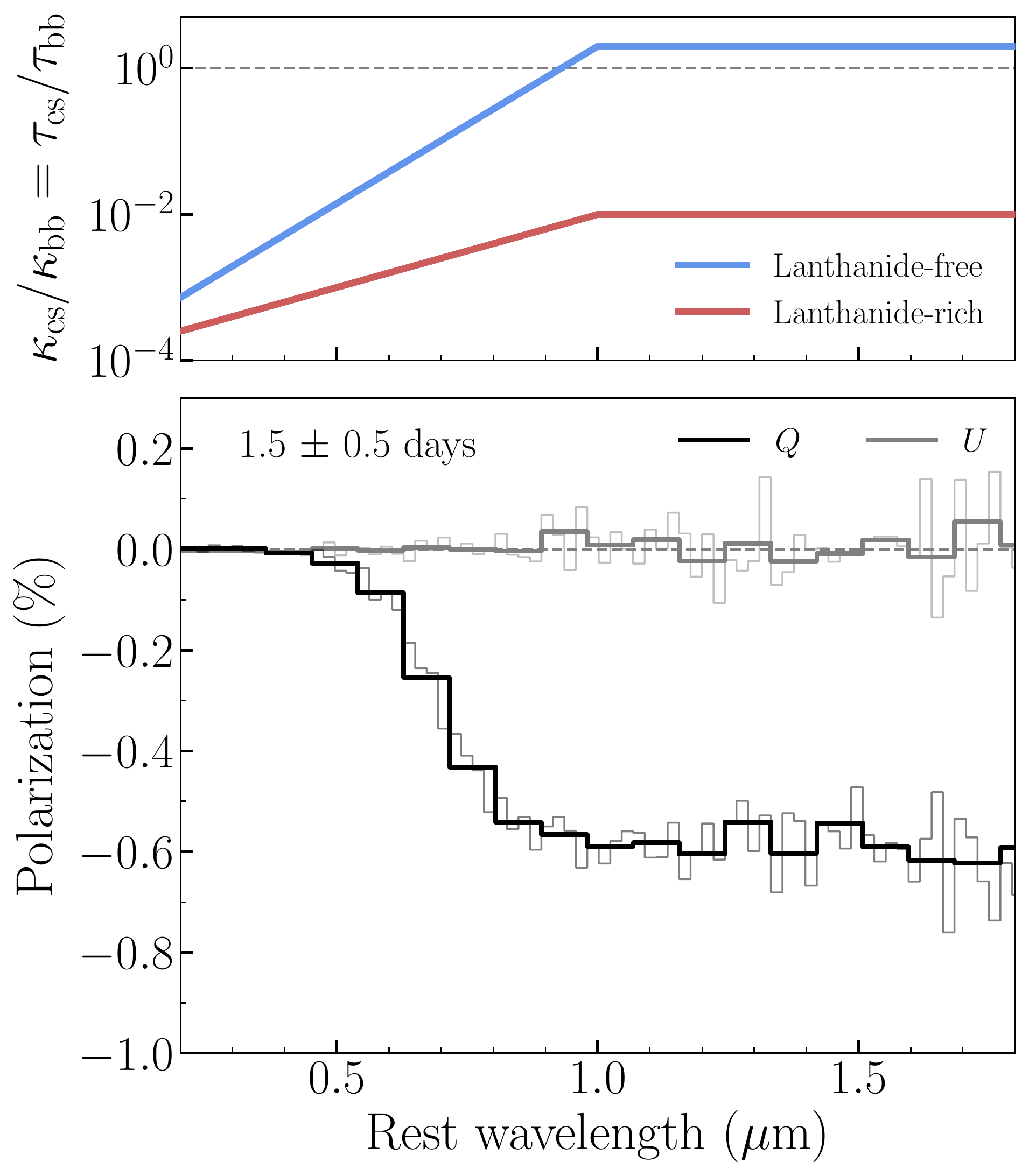}
\caption{Upper panel: relative importance of electron-scattering compared to bound-bound opacity ($\kappa_\mathrm{es}/\kappa_\mathrm{bb}$) at 1.5~d after the merger and at different wavelengths. Opacities in the lanthanide-free component are shown in blue, while those from the lanthanide-rich component in red. Lower panel: $Q$ (black) and $U$ (grey) polarization spectra at 1.5~$\pm$~0.5~d after the merger (average of 3 time-bins). Spectra as calculated from \textsc{possis} are shown with thin lines, while thick lines show a re-binned version to decrease the Monte Carlo noise (bin size = 4).}
\label{fig:pol}
\end{center}
\end{figure}

Polarization spectra at 1.5~d after the merger are shown in the bottom panel of Fig.~\ref{fig:pol}. Predictions refer to an equatorial orientation ($\cos\theta_\mathrm{obs}=0$), for which the polarization signal is expected to be maximized \citep{Bulla2019}. Moving the observer from the equator to the pole leads to a smaller polarization signal, with both $Q$ and $U$ consistent with zero when the system is viewed face-on ($\cos\theta_\mathrm{obs}=1$) due to the axial symmetry of the adopted model.

Given the axial symmetry of the model, the $U$ Stokes parameter is consistent with zero at all wavelengths. Following \cite{bulla2016}, deviations of $U$ from zero can thus be used as a proxy for Monte Carlo noise, which for the case of $N_\mathrm{ph}=2\times10^7$ used in these simulations is $\sigma = |U|\lesssim0.1$~per cent at all wavelengths. A net polarization signal is instead predicted across the $Q$ Stokes parameter. In line with what was found by \cite{Bulla2019}, all the polarization signal is created in the lanthanide-free region as electron scattering is a sub-dominant source of opacity in the lanthanide-rich component at all wavelengths ($\kappa_\mathrm{es}/\kappa_\mathrm{bb}\lesssim0.01$, see upper panel of Fig.~\ref{fig:pol}). The overall Stokes vectors coming from the lanthanide-free component are aligned in the horizontal direction, thus resulting in a negative $Q$ value (see also fig. 2 in \citealt{Bulla2019}).

The wavelength-dependence of the signal can be readily understood from the relative importance of electron scattering over bound-bound opacity ($\kappa_\mathrm{es}/\kappa_\mathrm{bb}$) in different spectral regions (see upper panel of Fig.~\ref{fig:pol}). At wavelengths bluer than 0.5~$\mu$m, $\kappa_\mathrm{es}/\kappa_\mathrm{bb}\lesssim0.01$ and thus the depolarizing effect of line opacities leads to $Q\sim0$. Moving from 0.5 to 1~$\mu$m increases $\kappa_\mathrm{es}/\kappa_\mathrm{bb}$ from $\sim$~0.01 to 2 (see equations~\ref{eq:kappaes} and \ref{eq:kappabb}), with the effect of increasing $Q$ from zero to $-$0.6~per cent. The same polarization level is finally predicted at all wavelengths larger than 1~$\mu$m, following the adopted choice of keeping the bound-bound opacity in the infrared fixed to the same value (see Section~\ref{sec:testmodel}).

The polarization signal drops very rapidly with time as a consequence of the fast increase of bound-bound opacity (i.e. decrease of $\kappa_\mathrm{es}/\kappa_\mathrm{bb}$, see Fig.~\ref{fig:opac}). $Q$ reaches values of $-0.2$~per cent in the infrared at 2.5~d after the merger and becomes negligible at later epochs. This behaviour is in good agreement with what was found in \cite{Bulla2019}, with differences in the absolute polarization levels due to the different choices for the opacities.

\section{Conclusions}
\label{sec:conclusions}

\referee{In this study, we presented \textsc{possis}, a Monte Carlo radiative transfer code that is well-suited to predict viewing-angle dependent observables for multi-dimensional models of SNe and KNe.} Building on previous works \citep{Bulla2015,Bulla2019}, we upgraded the code to incorporate an energy treatment of radiation and a time-dependence of both opacities and ejecta properties. Thanks to these upgrades, \textsc{possis} can calculate (i) spectral energy distributions (SEDs) at different times, (ii) broad-band light curves and (iii) polarization spectra for SN and KN models.

We tested \textsc{possis} against the two-component KN model discussed in \cite{Bulla2019}, in which the ejecta are characterized by a first component around the equatorial plane and rich in lanthanide elements and by a second component at polar regions and devoid of lanthanides. We presented synthetic observables for different viewing angles and demonstrated the power of \textsc{possis} to constrain the system inclination through the comparison of predicted SEDs, light curves and polarization spectra with KN observations.

Given the relatively fast computation times ($\sim$~hours on a single core for $N_\mathrm{ph}=10^5$ and $N_\mathrm{obs}=11$), \textsc{possis} \referee{using the abs-mode (see Section~\ref{sec:opac})} is well-suited to undertake parameter-space study to place constraints on key properties of SNe and KNe (e.g. ejecta mass, temperature, angular extent of the two components). Here, we presented a proof-of-concept of such parameter-space study by investigating the impact of the chosen ejecta mass and angular extent of the two components on the synthetic observables.

Although we focused on testing \textsc{possis} against a model with an idealized ejecta morphology, the code is completely flexible in terms of the input geometry. This will allow us to explore the more complex ejecta structure produced by multi-dimensional hydrodynamical models, predicting viewing-angle dependent observables that can be used to interpret data and place constraints on models.

\section*{Acknowledgements}
\referee{I thank the anonymous referee for helping to improve the quality of the paper.} I am very grateful to S. A. Sim, A. Goobar and H. F. Stevance for useful comments and suggestions. I acknowledge support from the G.R.E.A.T research environment funded by the Swedish National Science Foundation.

\bibliographystyle{mn2e}

{\footnotesize
\bibliography{bulla2019b}}

\end{document}